\documentclass[switch]{article}

\usepackage{preprint}

\usepackage{graphicx}
\usepackage{dcolumn}
\usepackage{bm}

\usepackage[utf8]{inputenc}
\usepackage[T1]{fontenc}
\usepackage{siunitx}
\usepackage{xspace}
\usepackage{braket}
\usepackage{sidecap}
\usepackage{float}

\newcommand*{\fig}{FIG.~}
\newcommand*{\sect}{Sect.~}

\newcommand{\ie}{i.e.\ }

\newcommand{\afmot}{AF-MOT\xspace}
\newcommand{\cwmot}{cw-MOT\xspace}

\newcommand{\hub}{Institut für Physik, Humboldt-Universität zu Berlin, 12489 Berlin, Germany}
\newcommand{\fbh}{Ferdinand-Braun-Institut, Leibniz-Institut für Höchstfrequenztechnik, 12489 Berlin, Germany}
\newcommand{\onlinecite}[1]{\hspace{-1 ex} \nocite{#1}\citenum{#1}}

\usepackage{amsmath, amsthm, amssymb, amsfonts}

\usepackage[numbers,square]{natbib}
\usepackage[utf8]{inputenc}	
\usepackage[T1]{fontenc}	
\usepackage{xcolor}		
\usepackage[colorlinks = true,
            linkcolor = purple,
            urlcolor  = blue,
            citecolor = cyan,
            anchorcolor = black]{hyperref}	
\usepackage{microtype}		
\usepackage{lineno}		
\usepackage{float}			
\usepackage{multicol}		

\usepackage{newfloat}
\DeclareFloatingEnvironment[name={Supplementary Figure}]{suppfigure}
\usepackage{sidecap}
\sidecaptionvpos{figure}{c}

\usepackage{titlesec}
\titlespacing\section{0pt}{12pt plus 3pt minus 3pt}{1pt plus 1pt minus 1pt}
\titlespacing\subsection{0pt}{10pt plus 3pt minus 3pt}{1pt plus 1pt minus 1pt}
\titlespacing\subsubsection{0pt}{8pt plus 3pt minus 3pt}{1pt plus 1pt minus 1pt}

\title{A single-laser alternating-frequency magneto-optical trap}

\usepackage{authblk}

\author[1\thanks{\tt{benjamin.wiegand@physik.hu-berlin.de}}]{B. Wiegand}
\author[1]{B. Leykauf}
\author[1]{K. Döringshoff}
\author[1]{Y. D. Gupta}
\author[1,2]{A. Peters}
\author[1,2]{M. Krutzik}

\affil[1]{\hub}
\affil[2]{\fbh}

\begin{document}
  
\maketitle

\begin{abstract}
In this paper, we present a technique for magneto-optical cooling and trapping of neutral atoms using a single laser. The alternating-frequency magneto-optical trap (\afmot) uses
an agile light source that sequentially switches between cooling and repumping transition frequencies
by tuning the injection current of the laser diode. We report on the experimental demonstration of such a system for \textsuperscript{87}Rb and \textsuperscript{85}Rb based on a micro-integrated extended cavity diode laser (ECDL) performing laser frequency jumps of up to \SI{6.6}{\giga\hertz} with a tuning time in the \SI{}{\micro\second} regime and a repetition rate of up to \SI{7.6}{\kilo\hertz}. For that, a combination of a feed-forward for coarse frequency control and a feedback for precise locking was used. We discuss the results of the \afmot characterization in terms of atom numbers and cloud temperature for different operation parameters.
\end{abstract}
\vspace{0.35cm}

\begin{multicols}{2} 


\section{Introduction}
The development of the magneto-optical trap\cite{magneto-optical-trap} (MOT) revolutionized the field of cold atom physics, providing a reliable technique for the production of cold and ultracold atomic clouds.
Recently, progress has been made in the development of novel miniaturized MOT geometries\cite{lee-pyramid-96,williamson-pyramid-1998,pyramid-mot,bouyer-pyramid-10,pollock-pyramid-11,mueller-pyramid-17,grating2,grating-imhof-17,grating-mot} which may enable the use of cold atom techniques even on challenging experimental platforms that put high demands on size, weight and power consumption such as those
in drop towers\cite{ai-quantus,degenerate-quantum-gases},
airplanes \cite{airplanes}
and small satellites \cite{nanosatellites}.
For further miniaturization of cold atom experiments, the laser system plays a crucial role as each laser requires driving and control electronics, temperature stabilization and light distribution hardware. Therefore, techniques for reducing the number of lasers are worth exploring.
A well-known approach for this involves laser modulation techniques for the generation of sidebands at repumping frequencies such that only a single laser is required\cite{sideband_mot,aom_mot}.

As an alternative approach, we present a novel technique for the operation of a MOT with a single laser that does not require optical modulators: the single-laser alternating-frequency magneto-optical trap (\afmot) uses an agile light source that sequentially targets cooling and repumping transitions by tuning the frequency of the laser.
This technique is applicable to species with moderate losses to dark states (\ie long lifetimes in the cooling cycle compared to the timescale of frequency tuning)
and is based on previous work performed at Leibniz University Hannover\cite{bartosch}.

This paper is structured as follows:
\sect\ref{sec:principle} introduces the principle of the \afmot.
\sect\ref{sec:setup} presents the experimental setup and \sect\ref{sec:experimental-sequence} explains our technique for the generation of frequency jumps between cooling and repumping transitions with a tuning time in the \SI{}{\micro\second} range.
In \sect\ref{sec:results} we show the results of our \afmot experiments with respect to atom number and temperature of the cloud.
In \sect\ref{sec:conclusion} we discuss our results.

\section{Principle of the \afmot}\label{sec:principle}
For laser cooling of \textsuperscript{87}Rb, we use cooling light that is slightly red-detuned from the $\ket{F=2}$ $\rightarrow$ $\ket{F'=3}$ transition of the $\text{D}_2$ line. Without repumping, this results in de-pumping
from the bright $\ket{F=2}$ to the dark $\ket{F=1}$ ground state
and a lifetime in the cooling cycle on the order of \SI{100}{\micro\second}. Repumping takes place much faster, on a timescale in the low \SI{}{\micro\second} regime\cite{mot-scaling-law}. For the operation of a MOT this means that short repumping pulses with a repetition rate in the \SI{}{\kilo\hertz} regime are sufficient for keeping a large fraction of the atoms in a bright state.

The \afmot principle follows from this idea and comprises a single laser that sequentially jumps between cooling and repumping transitions. One cycle of such an \afmot sequence 
\begin{figure}[H]
	\includegraphics[width=\linewidth]{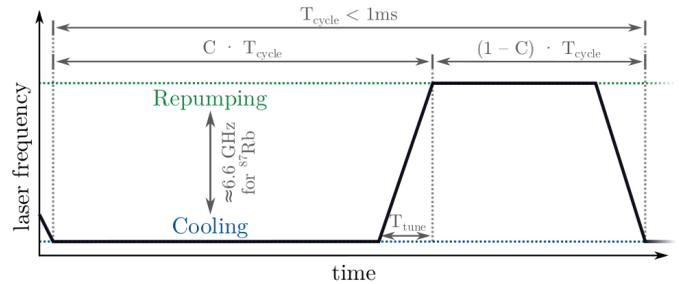}
	\caption[]{
	Illustration of the \afmot technique, showing a single cycle of the repetitive cooling and repumping sequence: the laser frequency (solid black) continuously alternates between cooling and repumping transitions (dashed blue and green, respectively), addressing each transition sequentially. \(T_\text{cycle}\) denotes the duration of such an \afmot cycle and $f_\text{cycle}=T_\text{cycle}^{-1}$ is the corresponding repetition rate; \(C\) denotes the percentage of time the laser frequency lock targets the cooling transition, including the time of laser frequency tuning \(T_\text{tune}\).
	}
	\label{fig:smot:sequence}
\end{figure}

\begin{figure}[H]
	\centering
	\includegraphics[width=.7\linewidth]{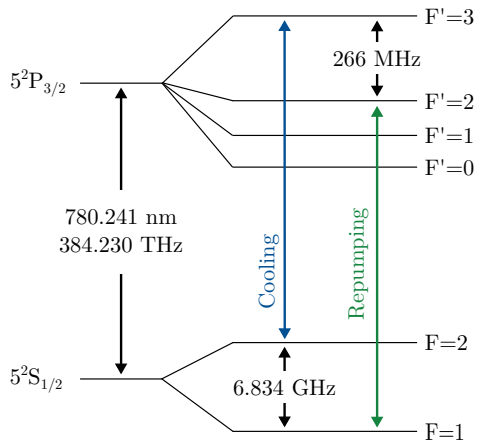}
	\caption[]{Level scheme of \textsuperscript{87}Rb with cooling and repumping transitions marked in blue and green, respectively. The frequency difference between these transitions is mainly given by the ground-state splitting and evaluates to roughly \SI{6.6}{\giga\hertz}.
	}
	\label{fig:rb:scheme}
\end{figure}

is illustrated in \fig\ref{fig:smot:sequence}: initially, the laser emits cooling light and thus captures atoms in the trap, while a fraction of the atoms decays into the dark state.
Then, a frequency jump is performed that targets the repumping transition, increasing the population of the bright state again. We use $C$ to label the fraction of time the laser targets the cooling transition (\ie the duty cycle of cooling light) including the tuning time of the laser $T_\text{tune}$.

For \textsuperscript{87}Rb, frequency jumps with an amplitude of \SI{6.6}{\giga\hertz} (see the level scheme in \fig\ref{fig:rb:scheme}) are necessary to sequentially address the cooling and repumping transitions. These frequency jumps have to be performed at a rate $f_\text{cycle}$ of more than \SI{\sim1}{\kilo\hertz} to ensure that a large fraction of the atoms remains in the bright state over multiple cycles. This requirement implies the need for fast frequency jumps, with tuning times $T_\text{tune}$ in the \SI{}{\micro\second} regime to maximize the time the laser frequency is close to an atomic transition.

\begin{figure*}[]
	\includegraphics[width=1.0\linewidth]{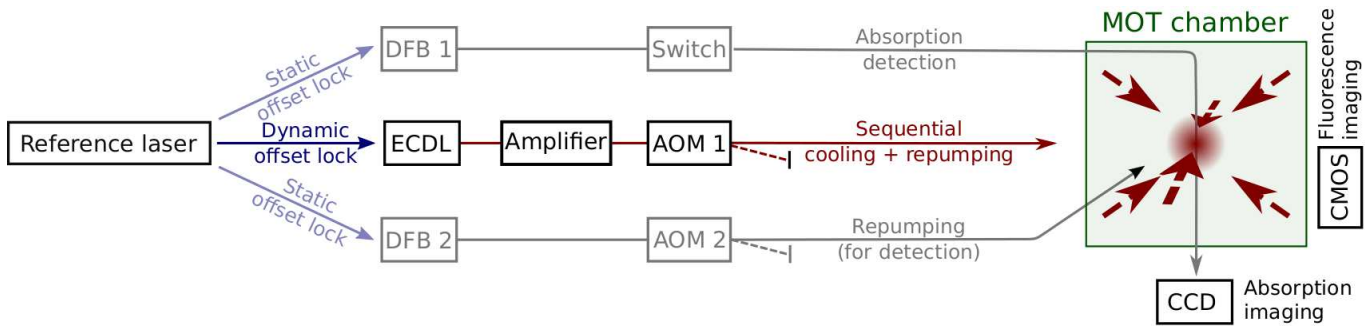}
	\caption{Schematic of the setup for \afmot experiments. For single-laser operation, only the central beam is relevant; the distributed feedback lasers (DFB 1 and 2) are merely required for absorption and fluorescence detection as well as for preparatory experiments. A charge coupled device (CCD) and a complementary metal-oxide-semiconductor (CMOS) camera are used to image the atomic cloud.}
	\label{fig:total_sketch}
\end{figure*}

\section{Experimental setup}\label{sec:setup}
For our experiments we use the MOT chamber of the Gravimetric Atom Interferometer GAIN \cite{gain-setup}. The trap loads atoms from the background vapor (\SI{5e-10}{\hecto\pascal}) and is formed by six beams in a 1-1-1 configuration, each with a light power of \SI{20}{\milli\watt} and an $\text{e}^{-2}$ diameter of $\SI{30}{\milli\meter}$. Two coils in anti-Helmholtz configuration produce a magnetic field with a central gradient of $\SI{0.5}{\milli\tesla\per\centi\meter}$.

The optical setup for the \afmot experiments is sketched in \fig\ref{fig:total_sketch}. A micro-integrated extended-cavity diode laser (ECDL) in a master-oscillator power-amplifier design based on the MiLas technology platform \cite{fbh-3,fbh-4} served as the light source for our \afmot experiments. It delivers light power of up to \SI{500}{\milli\watt} in-fiber and features a linewidth of \SI{100}{\kilo\hertz} (on a timescale of \SI{100}{\micro\second}). For the purpose of an \afmot, the single laser was dynamically offset locked (detailed in the next section) to a custom-built reference laser that was itself stabilized to the $\ket{F=2}$ $\rightarrow$ $\ket{F'=(1,2)}$ crossover transition of the \textsuperscript{87}Rb $\text{D}_2$ line by means of frequency modulation spectroscopy\cite{fms}. The resulting offset frequencies for cooling light (red-detuned to the transition by \SI{15}{\mega\hertz}) and resonant repumping light are given by \SI{251}{\mega\hertz} and \SI{6.845}{\giga\hertz}, respectively (including compensation for a frequency shift of $\SI{80}{\mega\hertz}$ introduced by AOM 1).

For technical reasons, the output of the ECDL was additionally amplified in the distribution module of the GAIN setup. We note that neither AOM 1, nor the two additional distributed feedback lasers (DFBs) are required for the \afmot technique itself; they were merely used for repumping the cloud in preparatory experiments as well as for fluorescence and absorption detection.

\section{Experimental sequence}\label{sec:experimental-sequence}
First, we prepare the laser system for frequency jumps on a \SI{}{\micro\second} timescale as required for the \afmot technique. For that, a feed-forward to the laser's injection current has to be generated. This task is performed by an iterative algorithm implemented on an FPGA that is described in detail in Sect. \ref{chap:preparation}. After several seconds, the algorithm converges and the generated feed-forward is suitable for driving fast frequency jumps. The algorithm continues to run, though, in order to compensate for laser drifts and to adapt the feed-forward signal to changing environmental conditions.

When we send the amplified laser light to the vacuum chamber, the MOT loading starts. When a steady-state is reached after a loading time of several seconds, we examine the atomic cloud by performing fluorescence and absorption measurements in order to determine the atom number and cloud temperature (Sect. \ref{chap:detection}).

\subsection{Generation of laser frequency jumps with a microsecond tuning time}
\label{chap:preparation}
\begin{figure*}[]
	\centering
	\includegraphics[width=1.0\linewidth]{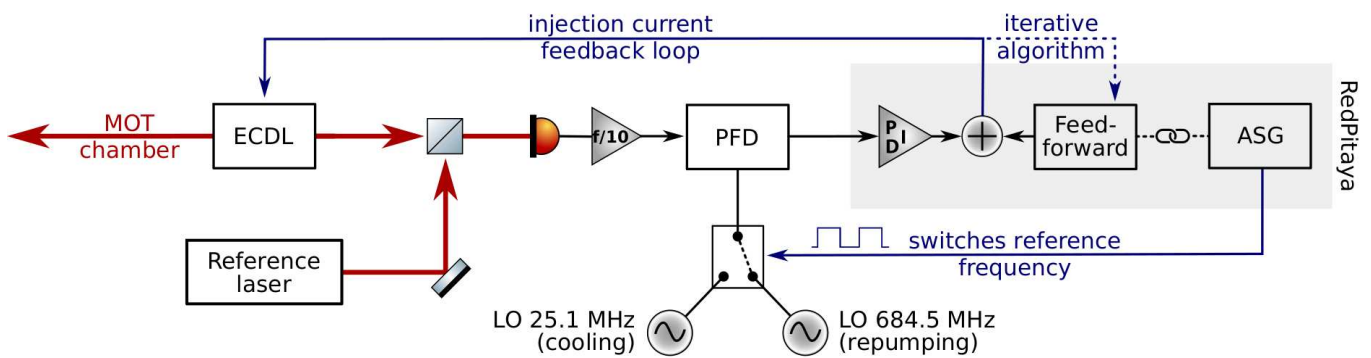}
	\caption[]{Schematic of the dynamic offset lock to generate the \afmot jump sequence. A phase-frequency detector compares the beat-note frequency with a reference frequency. An arbitrary sequence generator (ASG) generates a rectangular function that continuously alternates two oscillators that provide this reference; the PID servo thus aims at driving the laser frequency accordingly. Additionally, a synchronized feed-forward is added to the laser current. This feed-forward signal is generated by an FPGA algorithm before the actual experiments start; in order to compensate for laser frequency drifts, it is adapted continuously.}
	\label{fig:sketch}
\end{figure*}
For a \textsuperscript{87}Rb \afmot, frequency jumps of \SI{6.6}{\giga\hertz} with \SI{}{\mega\hertz}-accuracy have to be performed on a \SI{}{\micro\second} timescale. In order to simultaneously fulfill both requirements --- agility and precision ---
we chose a combined approach of\textbf{}
a feed-forward, for coarse frequency control, and
a feedback (PID), for precise locking. The setup used for this dynamic offset lock is depicted in \fig\ref{fig:sketch}.

The optical phase-locked loop (OPLL) consists of a radio frequency (RF) $f/10$ prescaler (RF Bay FPS-10-12) and a phase-frequency detector (PFD, OnSemiconductor MC100EP140) that compares the beat-note between ECDL and reference laser with an alternating RF reference frequency derived from two local oscillators (LO); the PID servo filter is implemented on the FPGA of a RedPitaya STEMlab device.
Its output controls the injection current of the ECDL's master oscillator (MO) via the modulation port of the current driver (ILX Lightwave 3724 in high-bandwidth mode). In order to switch between two different target frequencies of the offset lock (\ie cooling and repumping frequencies), an RF switch (composed of two MiniCircuits ZASWA-2-50DR+ and a ZFSC-4-1-S+ combiner) is triggered repetitively using the arbitrary sequence generator (ASG) of the RedPitaya.

As mentioned above, we found that we had to combine the OPLL's feedback signal to the laser injection current with a feed-forward in order to achieve fast frequency jumps. The naive approach of using a simple step function for this purpose is bound to fail, though, as the ECDL exhibits a complex frequency response to changes in the current on these short time scales. While to our knowledge no analysis of such fast frequency tuning of ECDLs exists, a study of the behavior of DFBs on very short timescales is published in Ref.~\onlinecite{dfb-step-response}. These two laser designs are similar in that a step in the injection current influences the laser frequency in the same two ways\cite{using-diode-lasers}: first, it alters the refractive index of the laser medium which happens almost instantaneously; secondly, thermal effects continue to shift the frequency.

We solved the problem by generating a non-trivial feed-forward that accounts for the complex frequency response of the laser. For the preparation of this signal we use an iterative algorithm that is fully implemented on the FPGA using Migen\cite{migen}:
to begin with, a sequence of alternating LO frequencies is initialized, with repetition rate and duty cycle corresponding to the desired values of $f_\text{cycle}$ and $C$. The PID then aims to drive the laser between these two target frequencies.
In each step of the iterative algorithm, the output of the PFD (containing information on when the laser frequency is lower or higher than the target frequency) is analyzed over one cycle: for each point in time, the average detuning at later times is calculated. This information is used to modify the feed-forward for the next step. Futhermore, several filters are applied to this signal in order to suppress oscillations.
As each iteration takes less than \SI{1}{\milli\second}, the algorithm converges within several seconds and the required frequency jumps can be performed within approximately \SI{15}{\micro\second} with moderate overshooting and ringing, fulfilling the requirements for an \textsuperscript{87}Rb \afmot.

\begin{figure}[H]
	\centering
	\includegraphics[width=\linewidth]{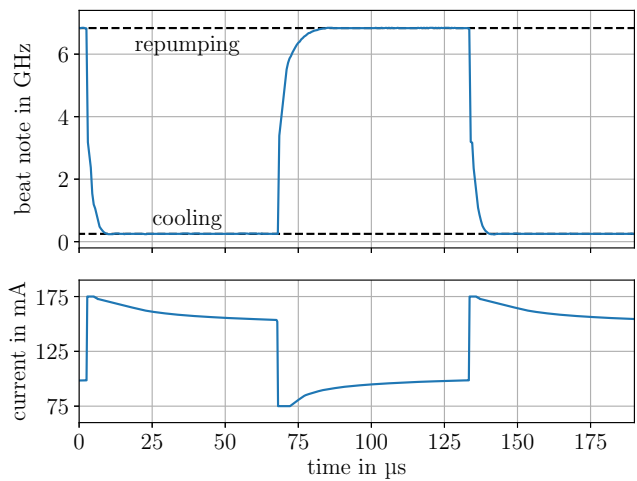}
	\caption[]{A record of the laser frequency and the MO's injection current for a single \afmot cycle (both plots share the same $x$-axis). The laser frequency continuously alternates between cooling and repumping frequencies of \textsuperscript{87}Rb (dashed black lines) with $f_\text{cycle}=\SI{7.6}{\kilo\hertz}$. For this plot, the laser's beat-note with a reference laser was recorded using a digital oscilloscope, and the time series was Fourier transformed. The injection current was calculated using the control signal of the servo loop and the tuning coefficient of the current driver. It was verified that the injection current indeed follows the modulation voltage with a phase delay of less than \SI{1}{\micro\second} even after steep changes.}
	\label{fig:jumps}
\end{figure}

An exemplary cycle of the \afmot sequence is depicted in \fig\ref{fig:jumps}.
The figure shows that the algorithm generates instantaneous current steps as large as \SI{100}{\milli\ampere} to perform the frequency jumps of \SI{6.6}{\giga\hertz}, although only roughly \SI{33}{\milli\ampere} are expected from the ECDL's slow tuning coefficient
(approximately $\SI{200}{\mega\hertz}/\SI{}{\milli\ampere}$).
This discrepancy is due to the aforementioned complex frequency response of the laser.
This is indicated in the graph, showing that subsequent to a frequency jump the feed-forward signal compensates for frequency drifts caused by temperature changes inside the laser.

\begin{figure}[H]
	\centering
	\includegraphics[width=\linewidth]{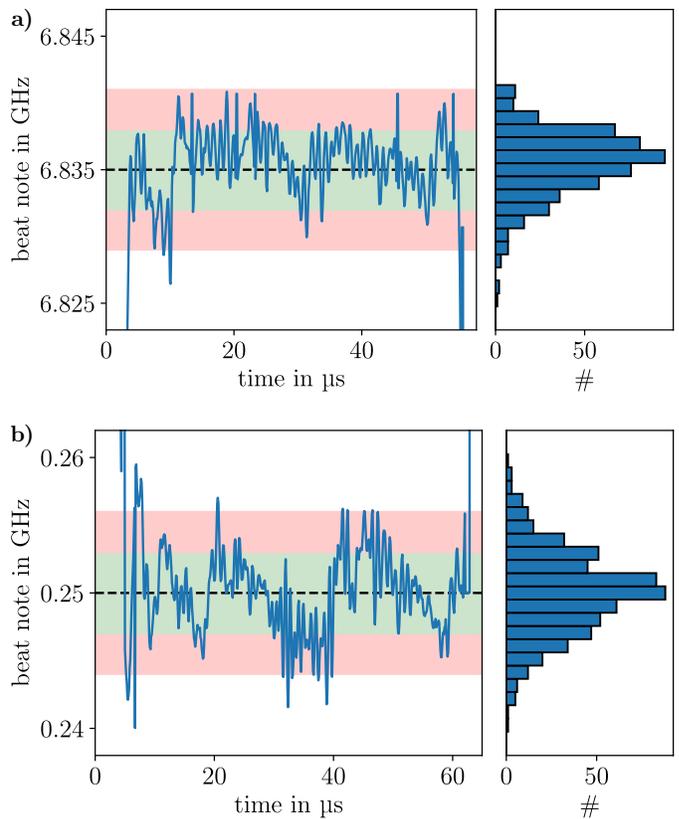}
	\caption[]{
		Part of a laser frequency jump sequence corresponding to the repumping (a) and cooling (b) plateau; histograms highlight the accuracy of the frequency tuning.
		The black dashed line depicts the target frequency, and the green and red regions visualize one or two times the linewidth of the \textsuperscript{87}Rb $\text{D}_2$ line (\SI{6}{\mega\hertz}), respectively.
		Each bar of the histogram represents a span of \SI{1}{\mega\hertz}.
		For this plot, the laser's beat-note was recorded with an oscilloscope and the time series was Fourier transformed using the short-time Fourier transform (STFT). Then, for each data point a sine function was fitted to the oscilloscope data, making use of STFT data as start parameters.
		This procedure allowed for a determination of the beat-note frequency with high sample rate and frequency resolution at the same time.
	}
	\label{fig:jump-detail}
\end{figure}

\fig\ref{fig:jump-detail} shows a zoom of the frequency jump cycle, highlighting the accuracy of the frequency tuning. In this plot, \SI{95}{}\,\% of the data points are within $\pm\Gamma$ around the target frequency, with $\Gamma=\SI{6}{\mega\hertz}$ being the linewidth of the \textsuperscript{87}Rb $\text{D}_2$ line; considering the whole cycle, including times of frequency tuning, this holds \SI{80}{}\,\% of the time.

The algorithm runs continuously during our experiments to compensate for laser drifts and to adapt the feed-forward signal to changing environmental conditions.
It is worth noting that the \afmot cycles for \textsuperscript{87}Rb were observed to run reliably for several hours without mode hops, despite the jump amplitude (\SI{6.6}{\giga\hertz}) being on the order of magnitude of the mode-hop-free tuning range of the ECDL (\SI{7}{\giga\hertz}).

We want to note that the method presented above is not specifically tailored for our ECDL and that it should be applicable to other laser sources such as commercially available macroscopic ECDLs, DFB or DBR laser sources as well. The major requirements are fast tunability between the required optical frequencies and a low frequency drift that may be compensated for by the slowly adapting feed-forward.
The use of a DFB or DBR laser as main light source may be advantageous as these laser designs typically exhibit a large mode-hop free tuning range and a higher tuning coefficient compared to ECDLs\cite{using-diode-lasers} that may allow for even faster frequency jumps. Additionally, it means that smaller injection current changes are sufficient for driving the \afmot sequence. Given that temperature changes cause mechanical stress, this may mitigate potential lifetime issues of the laser.

\subsection{Measurement of atom number and cloud temperature}
\label{chap:detection}
\begin{figure}[H]
	\centering
	\includegraphics[width=\linewidth]{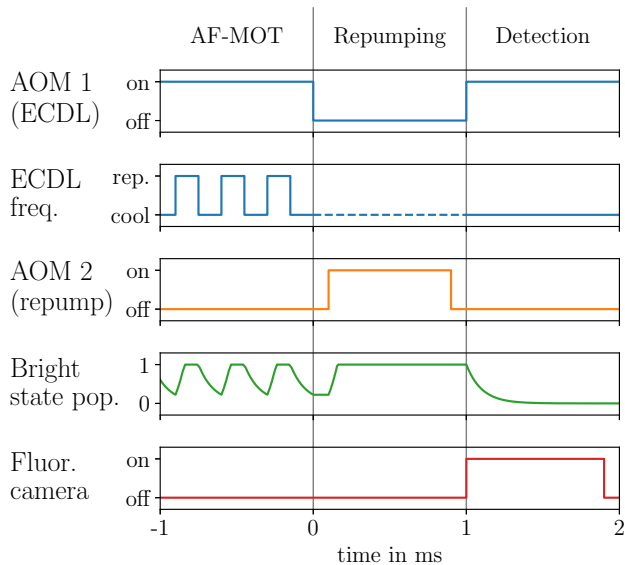}
	\caption[]{
		Schematic illustration of the detection scheme for measuring the atom number by means of fluorescence imaging. The first part shows \afmot operation with the ECDL alternating between cooling and repumping frequencies. As the fluctuation of bright state population is too fast for our camera, we cannot determine the atom number by analyzing the \afmot's fluorescence signal.
		Instead, we block the ECDL's output using AOM 1 and apply a repumping pulse of $\SI{.8}{\milli\second}$. Finally, we trigger the camera and unblock the ECDL (now locked to the cooling transition). As the camera's exposure time largely exceeds the time constant of decay to the dark state, we measure a signal that is proportional to the number of atoms inside the cloud.
		We note that the same measuring scheme is used for both MOT types (\afmot and \cwmot) to support comparability of the data.
	}
	\label{fig:detection-sequence}
\end{figure}

We use fluorescence imaging in order to determine the atom number of the MOT.
Images taken \emph{during} the \afmot sequence are not useful for these purposes, though, as the population of bright states varies over $T_\text{cycle}$ which we can not resolve given the minimum exposure time of our cameras. Instead, we implemented a procedure for repumping the cloud before taking an image of the fully repumped ensemble. The timing of this procedure is illustrated in \fig\ref{fig:detection-sequence}. After loading the MOT, the ECDL's output is blocked by AOM 1 and a dedicated repumping laser (DFB 2) is turned on using AOM 2, pumping all atoms to the bright state.

For fluorescence imaging, we then excite the atoms with cooling light and trigger the camera (The Imaging Source DMM 22BUC03-ML). In this way we record a fluorescence signal that is proportional to the number of atoms inside the cloud because the camera's exposure time of $\SI{.9}{\milli\second}$ is much longer than the time constant of decay to the dark state.

Additionally, we determine the temperature of the cloud by performing time-of-flight measurements using absorption imaging of the freely expanding atomic cloud. For that purpose, we switch off the magnetic field immediately after the the last AF-MOT cycle and block all laser light. We determine the optical density of the repumped cloud by absorption imaging (using a PCO Imaging Pixelfly 270 XS camera). To this end, an imaging beam (emitted by DFB 1) with a power of \SI{500}{\micro\watt} and an $\text{e}^{-2}$ diameter of \SI{30}{\milli\meter} is switched on by an optical switch (Agiltron NanoSpeed) and the camera is triggered with an exposure time of $\SI{5}{\micro\second}$.

\section{Results}\label{sec:results}
\subsection{Atom numbers for \textsuperscript{87}Rb}
In general, the number of atoms that can be trapped using the \afmot technique is lower than that of an equivalent \cwmot because the atoms are not continuously subjected to light forces. This results in a reduction of the capture velocity such that atoms from the high velocity tail of the temperature distribution cannot be trapped.

It is useful to characterize the \afmot atom number in terms of the atom number achieved in an equivalent steady-state \cwmot (which is \SI{7e8} atoms after \SI{10}{\second} in our set-up). We record this relative atom number as a percentage for a range of different switching cycles by varying the repetition rate $f_\text{cycle}$ and duty cycle $C$.
For an immediate comparison we sequentially load a conventional \cwmot and an \afmot for each data point to reject potential variations of the atom number on longer timescales.

\subsubsection{AOM-switched dual-laser \afmot}
\begin{figure*}[]
	\includegraphics[width=\linewidth]{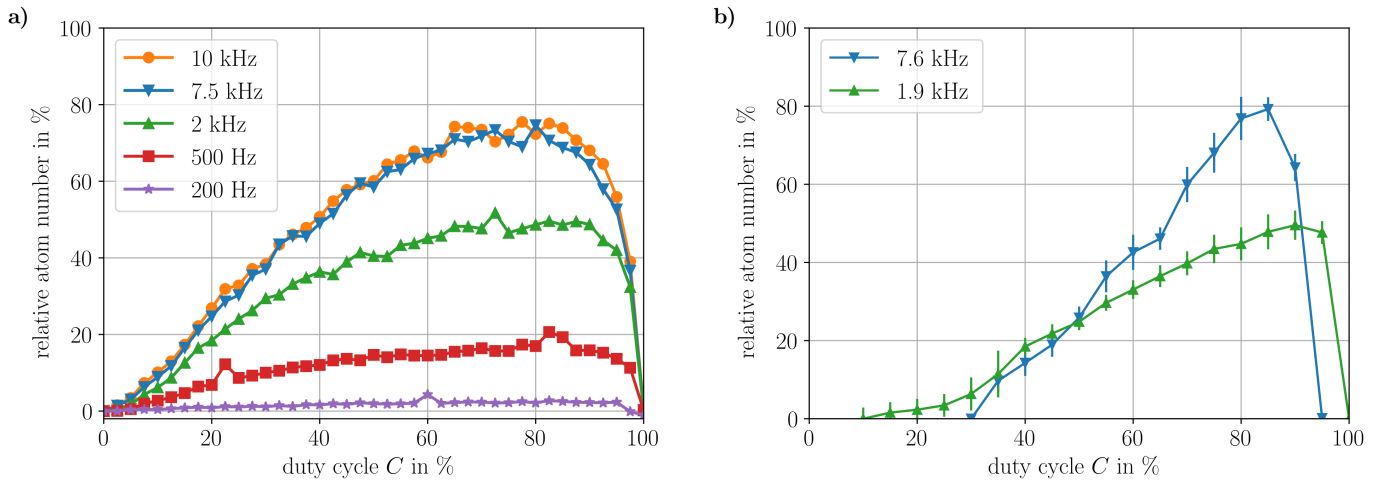}
	\caption{
		a)
		Relative steady-state atom number in a MOT when alternatingly switching cooling and repumping lasers on and off at different frequencies and duty cycles by means of AOMs. An equivalent conventional \cwmot with continuous cooling and repumping served as a basis of comparison.
		b)
		Relative steady-state atom numbers in an actual single-laser \afmot with $f_\text{cycle}=\SI{7.6}{\kilo\hertz}$; for each data point, 10 measurements were performed. For each duty cycle we sequentially measure the atom number of conventional \cwmot and \afmot to determine the relative atom number. We want to mention that a quantitative comparison of the atom numbers in a) and b) is only possible to a limited extent due to differences in laser power and unknown effects of finite AOM switching times.
	}
	\label{fig:fake-and-real-smot:atom-numbers}
\end{figure*}

In order to get an estimate for the maximum achievable performance of an ideal \afmot (\ie an \afmot without degradation due to finite tuning time and overshoot), we used a conventional \cwmot system and emulated an \afmot by alternately switching the cooling and repumping lasers using AOMs. The result is shown in \fig\ref{fig:fake-and-real-smot:atom-numbers}a. As the AOM switching time is in the \SI{}{\micro\second} regime, only repetition rates of up to \SI{10}{\kilo\hertz} were considered. We want to mention that this system is not perfectly equivalent to an actual single-laser \afmot as repumping is performed using an additional laser beam (DFB 2) with lower power from an extra port of the vacuum chamber, whereas the \afmot is repumped by the ECDL with strong pulses from all six MOT beams.

In agreement with our expectations, \afmot repetition rates in the \SI{}{\kilo\hertz} range (\ie cycle times of less than \SI{1}{\milli\second}) are required to reach atom numbers that are close to those of a conventional \cwmot. In this regime, faster repetition rates are advantageous as they lead to a higher bright state population and thus increase the average time an atom is subjected to light forces. Assuming perfect frequency tuning (\ie no dead time $T_\text{tune}$ between the different laser frequencies as well as no overshooting), relative atom numbers of $\SI{75}{\percent}$ are achievable with the \afmot technique for \textsuperscript{87}Rb and repetition rates of up to \SI{10}{\kilo\hertz}.

\subsubsection{Single-laser \afmot}
\label{chap:single-laser-afmot}
Having determined promising repetition rates and duty cycles by means of the AOM-switched dual-laser test system, we investigated the actual single-laser \afmot. Using the technique for dynamic offset locks described in Sect. \ref{chap:preparation}, continuous laser frequency jumps between cooling and repumping transitions of \textsuperscript{87}Rb were generated for different duty cycles $C$ with a repetition rate of \SI{7.6}{\kilo\hertz} (corresponding to a fraction of the FPGA's clock frequency). The resulting atom numbers are shown in \fig\ref{fig:fake-and-real-smot:atom-numbers}b. It should be noted that due to technical issues these experiments were conducted with a reduced laser light power, corresponding to roughly $2/3$ of the power used in the previous section for the AOM-switched dual-laser \afmot.

We find that the relative atom numbers of the single-laser \afmot roughly match the ones of the AOM-switched dual-laser \afmot. The main discrepancy is given by a reduced number of atoms for small and large duty cycles which we attribute to two effects: first, the finite laser frequency tuning time $T_\text{tune}$ of the single-laser \afmot corresponds to a dead time of $\SI{15}{\micro\second}$ during which the light may not excite atomic transitions, whereas cooling and repumping frequencies are alternated without delay in case of the AOM-switched \afmot.
Second, the relative impact of overshooting and ringing (apparent in \fig\ref{fig:jump-detail}) on the effective duty cycle $C$ or $1-C$ of the cooling and the repumping part, respectively, becomes more significant for shorter duty cycles.

Despite this difference at the edges of the plot, the maximum relative atom number achieved is roughly the same: the largest \afmot atom number was observed to be approximately \SI{5e8}{} for a duty cycle of $C=\SI{85}{\percent}$. With respect to a conventional \cwmot we found a relative atom number of approximately $\SI{75}{\percent}$ that is reached with about the same loading time as our \cwmot.

\subsection{Atom numbers for \textsuperscript{85}Rb}

As a proof of principle, the \afmot technique was also demonstrated using \textsuperscript{85}Rb. In this case, frequency jumps of \SI{2.9}{\giga\hertz} were required, corresponding to the difference between cooling ($\ket{F=3}$ $\rightarrow$ $\ket{F'=4}$) and repumping ($\ket{F=2}$ $\rightarrow$ $\ket{F'=3}$) transitions of the \textsuperscript{85}Rb $\text{D}_2$ line. This allows for faster frequency tuning times compared to \textsuperscript{87}Rb whose ground state splitting is roughly twice as large. On the other hand, the lighter isotope exhibits a smaller line splitting which leads to faster decay to the dark state. Experimentally, a relative atom number of \(\SI{35}{\percent}\) with respect to the case of a conventional dual-laser \textsuperscript{85}Rb \cwmot was achieved with a repetition rate of \SI{7.6}{\kilo\hertz} and a duty cycle of \(C=\SI{90}{\percent}\).

\subsection{Temperature for \textsuperscript{87}Rb}
When performing time-of-flight measurements by means of absorption imaging of a freely expanding cloud, we found no significant differences in temperature between \afmot and \cwmot ($\approx\SI{500}{\micro\kelvin}$).

\section{Conclusion and outlook}\label{sec:conclusion}
We have demonstrated a novel technique for single-laser magneto-optical cooling and trapping of neutral atoms and characterized the performance of such an \afmot for different operation parameters.

Preparatory experiments were conducted using a conventional MOT system in order to estimate the achievable performance with this technique. In general, the reduced exposure time to cooling light forces leads to lower capture velocities and thus a smaller cloud. Experimentally, we found that for \textsuperscript{87}Rb the required repetition rate of the cooler-repumper frequency jump sequence is higher than \SI{1}{\kilo\hertz}.
With an optimized duty cycle, this yields a relative atom number of roughly \SI{75}{\percent} compared to an equivalent conventional \cwmot.
We point out that the AOM-switched system that was used for these preparatory experiments is not completely equivalent to a single-laser \afmot due to a different repumper intensity and a lack of frequency tuning (\ie no laser light of intermediate frequencies influences the atoms). However, it yields a good estimate for the potential of the \afmot technique.

We have implemented a dynamic offset lock, capable of driving laser frequency jumps of $\SI{6.6}{\giga\hertz}$ (corresponding to the difference between cooling and repumping transitions of \textsuperscript{87}Rb) with a tuning time of approximately $\SI{15}{\micro\second}$. Finally, we used this technique to demonstrate a single-laser \afmot with a maximum atom number of \SI{5e8}{}, which is \SI{75}{\percent} of the equivalent \cwmot. As this atom number roughly corresponds to the one of the AOM-switched dual-laser \afmot, we conclude that our method for the generation of fast frequency jumps works in a satisfactory manner, reaching a tuning precision that has no substantial negative impact on the atom number.

So far, the dynamic offset lock employed by our \afmot setup requires a second laser, providing an absolute frequency reference. This means that sophisticated experiments that rely on such a reference for other reasons than cooling and trapping can directly profit from a reduced number of lasers.
Additionally, some bare-bone MOT experiments may take advantage of the \afmot technique as well. For cooling and trapping of atomic species that exhibit more than a single dark state, an agile laser could sequentially target several repumping frequencies. Moreover, the \afmot technique is not limited to the trapping of a single species; due to the similar level structure of \textsuperscript{85}Rb and \textsuperscript{87}Rb, even a single-laser dual-species \afmot, sequentially addressing 4 different transitions, is conceivable.

Actual single-laser operation (\ie self-referencing of the \afmot laser) may be feasible in future experiments. For that purpose, two different approaches are conceivable. First, a spectroscopy signal could be recorded periodically, either by making use of the fast frequency tuning during the \afmot jump sequence or by adding a dedicated ramp to the cycle. Alternatively, a feedback loop based on the fluorescence signal strength alone could be implemented as demonstrated in Ref.~\onlinecite{mot-lock}.

\footnotesize
\section*{Acknowledgements}
We thank Thijs Wendrichs for the detailed review of the manuscript and Victoria Henderson for proof-reading the manuscript and for her helpful comments. We are grateful to Robert Jördens for his advice on FPGA programming.

This work is supported by the German Space Agency (DLR) with funds provided by the Federal Ministry of Economics and Technology (BMWi) under grant numbers DLR50WM1857, DLR50WP1702 and DLR50WP1432.

\nocite{*}
\normalsize
\bibliography{ms}





\end{multicols}  
\end{document}